# A fully automatized method for the unambiguous wavelength-by-wavelength determination of the thickness and optical property of a very thin film with a transparent range

Florian Maudet[1,a)], Charlotte Van Dijck[1], Muhammad Hamid Raza[1], Catherine Dubourdieu[1,2]

[1]Helmholtz-Zentrum-Berlin für Materialien und Energy GmbH, Institute "Functional Oxides for Energy Efficient Information Technology", 14109 Berlin, Germany

[2]Freie Universität Berlin, Physical Chemistry, Arnimallee 22, 14195 Berlin, Germany

a) Electronic mail: maudet.florian@yahoo.fr

Abstract

Spectroscopic ellipsometry is a powerful method with high surface sensitivity that can be used to monitor the growth of even sub-monolayer film. However, the analysis of ultrathin films is complicated by the correlation of the dielectric constant and the thickness. This problem is usually resolved by fixing one or the other value, limiting the information that can be extracted. Here, we propose a method to determine unambiguously the refractive index, extinction coefficient and thickness of a film when a transparent range is available in the energy range investigated. We decompose the analysis in three steps. First, the thickness of the film is determined from the transparent range of the film. Then, knowing the thickness of the layer, an initial estimation of the refractive index and extinction coefficient is made based on a first-order Taylor expansion of the ellipsometric ratio. Finally, using this estimation, a numerical regression is done to ensure the convergence of the fit towards the solution. A theoretical example of the method is given for two different thicknesses of $TiO_2$ films. Finally, the method is applied to the experimental data measured during the atomic layer deposition of a thin film of $Hf_{0.5}Zr_{0.5}O_2$ grown on Si. The thickness, refractive index and extinction coefficient are retrieved with a high precision in the energy range of 3.5 – 6.5 eV. A detailed analysis is presented on the accuracy of the retrieved values and their dependency on random and systematic errors for different energy ranges.

## I. Introduction

Spectroscopic ellipsometry is an optical, non-destructive, characterization method commonly used to precisely monitor the growth of thin films both in research and industry.[1] This method relies on the measurement of the complex ellipsometric ratio, ρ, that characterize the changes in polarization after a linearly polarized light interacts with a sample. Remarkably the measurement is made without the need to calibrate the background intensity as opposed for example to spectrophotometry, an aspect that consequently enhance the reliability of the measurement.[2] Furthermore owing to its high sensitivity to surface change and low footprint, the method is particularly suited for in-situ study.[3] In the simple case of an isotropic three-phase configuration consisting of the ambient (with complex refractive index $n_a$), thin film ($n_f$) and substrate ($n_s$), ellipsometry allows to determine the unknown thickness $d_f$ and dielectric constant of the thin film $\varepsilon_f = (\tilde{n}_f)^2 = (n_f + ik_f)^2$ where $\tilde{n}_f$, $n_f$ and $k_f$ are the complex refractive index, real refractive index and extinction coefficient of the thin film respectively. To do so in most cases the retrieval of $d_f$ and $\tilde{n}_f$ from the measured ρ is made by developing an optical model assuming a certain dispersion property of the dielectric constant.[1] Indeed due to the non-linearity of the optical equations no direct inversion can be made.[4] The information is retrieved by varying the parameters of the optical model to minimize the Mean Square Error (MSE) that characterizes the error between the measured and modeled data. Therefore, a prior estimation of the optical properties is needed to ensure the convergence of the model towards a realistic solution. This method can lead to incorrect optical properties if spectral features of the film, that were not anticipated and therefore absent from the model, are overlooked. This is the case for example for a sample with sub-bandgap absorption features that would not have been taken into account with a simple Tauc-Lorentz model.[5] Furthermore, in the case of a very thin film ($\frac{d_f}{\lambda} \ll 1$ with λ the wavelength) this approach cannot be applied as $\tilde{n}_f$ and $d_f$ become strongly correlated.[2] This is particularly problematic for the very first steps of the growth of a thin film. This issue is usually overcome by fixing either $d_f$ or $\tilde{n}_f$. For example, to study an atomic layer deposited (ALD) film it is common to fix the optical property of a growing film to the bulk value and to recover information on the film growth from the thickness evolution.[6–9] However, in addition to preventing us from retrieving information on the dielectric constant of the film, this method leads to incorrect values of the thickness in the case of ultrathin films (typically below 10nm) as $\tilde{n}_f$ depends on the film thickness (an ultrathin film of e.g. 0.8 nm has a refractive index different from the one of the bulk).

A method to avoid this issue is to use complementary measurements to disambiguate $d_f$ and $\tilde{n}_f$, such as measuring the mass of the deposited material with a quartz crystal microbalance.[10] Another method was developed relying on the simultaneous measurement of changes in the reflected intensity and ρ to disambiguate $d_f$ and $\tilde{n}_f$.[11] A drawback is the need for a precise measurement of the intensity as it dominates the measurement accuracy.[11] Another approach solely relying on the measurement of ρ was developed by minimizing the presence of artefacts from the substrate in the dielectric constant for incorrect thicknesses.[12,13] This method allows to unambiguously determine the thickness when a substrate presents sharp feature, i.e. with a high variation with energy, like a critical point. An a priori knowledge of $\tilde{n}_f$ is, however, necessary to ensure the convergence towards the correct solution as multiple solutions of $\tilde{n}_f$ coexist for a given $d_f$.[4] Finally, when the material is transparent ($k_f = 0$), $\tilde{n}_f$ is a real at least on part of the investigated spectral range, a direct inversion of the thickness and refractive index can be made.[14] The method was extended recently to take into account the error on $d_f$, enhancing the accuracy.[15] Knowing $d_f$, $\tilde{n}_f$ can then be calculated for the whole spectral range by mapping all existing solutions and selecting the solution that is physically reasonable.[16] This step is computationally intensive and requires a manual selection of the solution to ensure that a physical solution is found. It limits the applicability for a real time analysis of a growing film. Therefore, a point-by-point method to unambiguously determine $d_f$ and $\tilde{n}_f$ without prior assumption on $\tilde{n}_f$ or manual selection of the solution is desirable to study the growth of very thin films *in situ* and in real time.

Here, we propose to address this issue by developing a fully automated method to determine unambiguously $d_f$, $n_f$ and $k_f$ of a very thin film where a transparent range is available. The method is well suited to study the growth of dielectrics or semiconductors with a bandgap that lies in the measured range. We decompose the analysis in three steps. First, the thickness of the film is determined following the procedure developed by F. L. McCrakin and M. Gilliot *et al.*.[14,15] Then, knowing the thickness of the layer, the refractive index and extinction coefficient can be retrieved for the whole spectral range without any prior knowledge on the film using a first order Taylor expansion of the ellipsometric ratio $\rho$ as proposed by G.H. Jung et al.[17] Finally, using this first order approximation to ensure convergence toward the correct solution, a wavelength-by-wavelength regression is made. A theoretical example is presented for a thin film of TiO$_2$. A detailed analysis and discussion on the error of $d_f$, $n_f$ and $k_f$ for different thicknesses of TiO$_2$ is presented. Finally, the method is demonstrated for the practical example of a very thin film of Hf$_{0.5}$Zr$_{0.5}$O$_2$ (HZO) grown by ALD.

## II. Model and method to unambiguously determine $\tilde{n}_f$ and $d_f$

The method can be divided in three parts: a first part to determine the thickness of the film, a second one to have an estimation of the optical properties from the calculated thickness and a third one that uses this estimation to ensure convergence towards the actual solution by a numerical regression. We consider here the simple case of an isotropic three-layer configuration: ambient ($\tilde{n}_a$), thin film ($\tilde{n}_f$) and substrate ($\tilde{n}_s$) as illustrated on Figure 1.

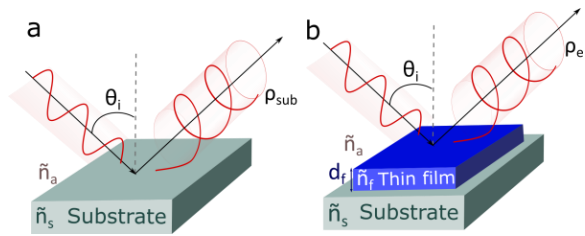

**Figure 1:** Schematic of an ellipsometry measurement on a bare substrate (a) and in three-phase ambient/thin film/substrate configuration (b)

**A. Thickness determination from a McCrakin inversion:**

We remind below the procedure that was originally presented by F. L. McCrakin to evaluate the thickness of a transparent thin film and further developed recently.[14,16] As we are considering a transparent thin film, $\tilde{n}_f = n_f$. The experimentally measured ellipsometric ratio is given by:[1]

$$\rho_e = \frac{r_p}{r_s} = \tan \psi_e \, e^{-i\Delta_e} \quad (1)$$

where $\psi_e$ and $\Delta_e$ are the measured ellipsometric angles, $r_p$ and $r_s$ are the p and s polarized complex reflection coefficients of the stack respectively. These coefficients are given by:

$$r_p = \frac{r_{af,p} + r_{fsub,p} X}{1 + r_{af,p} r_{fsub,p} X} \text{ and } r_s = \frac{r_{af,s} + r_{fsub,s} X}{1 + r_{af,s} r_{fsub,s} X} \quad (2)$$

$$\text{with } X = e^{\frac{j 4\pi \tilde{d}_f \sqrt{\tilde{n}_f^2 - \tilde{n}_a^2 \sin^2 \theta_i}}{\lambda}} \quad (3)$$

where $r_{af,p(s)}$, $r_{fsub,p(s)}$ are the Fresnel reflection coefficients of the ambient/thin film interface and thin film/substrate interface of the p (s) polarization respectively. It should be noted that we are using the physical convention: $\tilde{n}_f = n_f + i k_f$ leading to the negative sign in Eq. (1) and its absence in Eq. (3).[1]

From Eq. (3) $\tilde{d}_f$ can be expressed as:

$$\tilde{d}_f = \frac{-j\lambda \ln(X) + 2m\lambda\pi}{4\pi \sqrt{n_f^2 - \tilde{n}_a^2 \sin^2 \theta_i}} \quad (4)$$

where $\theta_i$ is the angle of incidence and $m$ is an integer that takes into account the multiplicity of orders due to the periodic behavior of $X$. In the case of very thin films $m = 0$.

From there, the goal is to calculate $X$ in order to determine $\tilde{d}_f(n_f)$ from Eq. (4). From Eq. (1) and Eq. (2) it follows that:

$$\rho_e = \frac{\frac{r_{af,p} + r_{fsub,p} X}{1 + r_{af,p} r_{fsub,p} X}}{\frac{r_{af,s} + r_{fsub,s} X}{1 + r_{af,s} r_{fsub,s} X}} = \frac{r_{af,p} + r_{af,p} r_{af,s} r_{fsub,s} X + r_{fsub,p} X + r_{fsub,p} r_{af,s} r_{fsub,s} X^2}{r_{af,s} + r_{fsub,s} X + r_{af,p} r_{fsub,p} r_{af,s} X + r_{af,p} r_{fsub,p} r_{fsub,s} X^2} \quad (5)$$

$$(\rho_e r_{af,p} r_{fsub,p} r_{fsub,s} - r_{fsub,p} r_{af,s} r_{fsub,s}) X^2 + (\rho_e r_{fsub,s} + \rho r_{af,p} r_{fsub,p} r_{af,s} - r_{af,p} - r_{fsub,p}) X + \rho_e r_{af,s} - r_{af,p} = 0$$
(6)

From Eq. (6), for a given $n_f$ two solutions of $X$ can be found. Consequently, using those solutions two values of $\tilde{d}_f(n_f)$ can be calculated. The main idea of the procedure is that, as the thickness must be a real number, the correct value of $n_f$ is the one that cancels out the imaginary part of the thickness, such that:

$$Im(\tilde{d}_f(n_f)) = 0 \quad (7)$$

These two values of $\tilde{d}_f(n_f)$ can be numerically computed for the whole wavelength (or energy) range with the method described in [16]. To do so, the values of $\tilde{d}_f$ are computed for a broad range of $n_f$ values, typically $n_f = [1 - 10]$ with 100 steps. Approximated values of $n_f$ are given by those values corresponding to the change in the sign of the imaginary part of $\tilde{d}_f$. From these initial approximations, the precise values of $n_f$ are then finally computed using an algorithm to find the root of Eq. (7). The algorithm used in this work is the Newton-Raphson method [18]. The sign ambiguity in the solution of Eq. (6) is solved by keeping the solution that makes physical sense ($d_f > 0$). As this computation is made for all the measured wavelengths, $d_f$ can be presented as a function of energy ($E$). Although the measurements are usually made with a fixed wavelength step, the results will be discussed as a function of energy since this scale is more relevant to discuss the material properties like the bandgap. Obviously $d_f$ should be constant for all energies. However, two causes can explain the energy dependency of $d_f$:

- This computation is made with the hypothesis that $k_f = 0$, therefore $d_f$ will vary in the energy range where this is not true.

- Measurements inevitably contain errors which cause variation of $d_f$ with the energy. Evaluating the impact of the errors of the measurement on $d_f$ is thus critical to evaluate the range of energy where it can be accurately determined.[15]

The error of $d_f$ can be calculated from the propagation error formula as follows [19]:

$$\sigma_{d_f} = \sqrt{\begin{array}{l}\left(\frac{\partial d_f}{\partial \psi}\right)^2 \sigma_\psi^2 + \left(\frac{\partial d_f}{\partial \Delta}\right)^2 \sigma_\Delta^2 + \left(\frac{\partial d_f}{\partial \theta_i}\right)^2 \sigma_{\theta_i}^2 + \left(\frac{\partial d_f}{\partial n_{sub}}\right)^2 \sigma_{n_{sub}}^2 + \left(\frac{\partial d_f}{\partial k_{sub}}\right)^2 \sigma_{k_{sub}}^2 + \left(\frac{\partial d_f}{\partial \lambda}\right)^2 \sigma_\lambda^2 \\ + 2\left(\frac{\partial d_f}{\partial \psi}\right)\left(\frac{\partial d_f}{\partial n_{sub}}\right)\sigma_{\psi n_{sub}}^2 + 2\left(\frac{\partial d_f}{\partial \Delta}\right)\left(\frac{\partial d_f}{\partial n_{sub}}\right)\sigma_{\Delta n_{sub}}^2 \\ + 2\left(\frac{\partial d_f}{\partial \psi}\right)\left(\frac{\partial d_f}{\partial k_{sub}}\right)\sigma_{\psi k_{sub}}^2 + 2\left(\frac{\partial d_f}{\partial \Delta}\right)\left(\frac{\partial d_f}{\partial k_{sub}}\right)\sigma_{\Delta k_{sub}}^2 \\ + 2\left(\frac{\partial d_f}{\partial \psi}\right)\left(\frac{\partial d_f}{\partial \theta_i}\right)\sigma_{\psi \theta_i}^2 + 2\left(\frac{\partial d_f}{\partial \Delta}\right)\left(\frac{\partial d_f}{\partial \theta_i}\right)\sigma_{\Delta \theta_i}^2 \\ + 2\left(\frac{\partial d_f}{\partial \psi}\right)\left(\frac{\partial d_f}{\partial \lambda}\right)\sigma_{\psi \lambda} + 2\left(\frac{\partial d_f}{\partial \Delta}\right)\left(\frac{\partial d_f}{\partial \Delta}\right)\sigma_{\Delta \lambda}\end{array}} \quad (8)$$

where the $\sigma_j$ is the standard deviation of the associated parameter $j$ and $\sigma_{xy}^2$ is the covariance of the parameters $x, y$. Therefore, by looking at the energy range where $\sigma_{d_f}$ and $\left|\frac{\partial d}{\partial E}\right|$ are minimum, the thickness can be accurately evaluated. In practice, the thickness of the film is determined as a weighted average of $d_f(E)$ with the weights $w(E)$. The weights are calculated to minimize the values of $\sigma_{d_f}$ and $\left|\frac{\partial d}{\partial E}\right|$ using the following function:

$$w(E) = \frac{1}{\left|\frac{\partial d_f(E)}{\partial E}\right|\sigma_{d_f}(E)} \quad (9)$$

Although it was presented here for the case of a three-phase configuration it can also be applied for a multi-layer stack where one layer is unknown. The only restriction of this method is that the thin film should exhibit a transparency range within the measurement range.[16]

### B. Determination of $\tilde{n}_f$ from first order Taylor expansion

Knowing $d_f$ is not sufficient to disambiguate $n_f$ and $k_f$ from ellipsometric measurements since multiple solutions of $\tilde{n}_f$ coexist for a given thickness.[4] A method was proposed recently by G.H. Jung et al. [17] to approximate $\tilde{n}_f$ without any *a priori* knowledge in the case of very thin films $\left(\frac{d_f}{\lambda} \ll 1\right)$. It relies on the first-order Taylor expansion of $\rho$. They evidenced that, in such a configuration, $\tilde{n}_f$ can be approximated by:

$$\tilde{n}_f^2 \approx \frac{1}{2}\left(\tilde{n}_a^2 + \tilde{n}_{sub}^2 + \frac{\delta_\rho}{\alpha}\right) \pm \frac{1}{2}\sqrt{\left(\tilde{n}_a^2 + \tilde{n}_{sub}^2 + \frac{\delta_\rho}{\alpha}\right)^2 - 4\tilde{n}_a^2\tilde{n}_{sub}^2} \quad (10)$$

$$\text{with, } \alpha = 4i\frac{2\pi}{\lambda}d_f\frac{\tilde{n}_a\tilde{n}_{sub}^2\cos(\theta_i)\sin^2(\theta_i)}{(\tilde{n}_a-\tilde{n}_{sub}^2)(\tilde{n}_{sub}^{2-\tilde{n}_a^2}+(\tilde{n}_a^2+\tilde{n}_{sub}^2)\cos(2\theta_i))} \quad (11)$$

$$\text{and } \delta_\rho = \frac{\rho_e-\rho_{sub}}{\rho_{sub}} \quad (12)$$

where $\rho_{sub}$ is the ellipsometric ratio of the substrate before thin film deposition (Figure 1(a)). It can be either measured before thin film deposition or simulated from the known $\tilde{n}_{sub}$. The ambiguity in the sign can be removed by choosing the solution that is the closest to the refractive index as determined from Eq. (7) in the transparent range of the film.

### C. Numerical regression using Newton-Raphson algorithm from the $\tilde{n}_f$ first order Taylor expansion

Since the aforementioned method to determine $\tilde{n}_f$ is a first order approximation, there will necessarily be a residual error between the modeled ellipsometric ratio $\rho_m$ (that can be calculated from Eq. (5)) and $\rho_e$. To minimize it, a regression can be made to refine the determination of $\tilde{n}_f$. The Newton-Raphson method can be used, as described here in matrix form for simplicity:

We define the initial vector $V_{f,0}$ as:

$$V_{f,0} = \begin{pmatrix} n_{f,0} \\ k_{f,0} \end{pmatrix} \quad (13)$$

where $n_{f,0}$ and $k_{f,0}$ are the values calculated from Eq. (10). Their values can then be refined in an iterative process by:

$$V_{f,j+1} = V_{f,j} + J_{df,j}^{-1} \cdot \Delta V_{d,j} \quad (14)$$

with the error vector $\Delta V_{d,j} = \begin{pmatrix} Re(\rho_e) - Re(\rho_m)_j \\ Im(\rho_e) - Im(\rho_m)_j \end{pmatrix} \quad (15)$

and the Jacobian matrix $J_{df} = \begin{pmatrix} \frac{\partial Re(\rho_m)}{\partial n} & \frac{\partial Re(\rho_m)}{\partial k} \\ \frac{\partial Im(\rho_m)}{\partial n} & \frac{\partial Im(\rho_m)}{\partial k} \end{pmatrix} \quad (16)$

The algorithm is repeated until convergence. At this step, $d_f$, $n_f$ and $k_f$ are determined unambiguously for a very thin film.

Finally, an important aspect of this method is to estimate the error $\sigma_{\tilde{n}_f}$ on $n_f$ and $k_f$ to be able to discriminate physical features of the spectra from a measurement artefact. We apply again the propagation error formula leading to the following error expression: [19]

$$\sigma_{\tilde{n}_f} = \sqrt{\begin{aligned} & \left(\frac{\partial \tilde{n}_f}{\partial \psi}\right)^2 \sigma_\psi^2 + \left(\frac{\partial \tilde{n}_f}{\partial \Delta}\right)^2 \sigma_\Delta^2 + \left(\frac{\partial \tilde{n}_f}{\partial d_f}\right)^2 \sigma_{d_f}^2 + \left(\frac{\partial \tilde{n}_f}{\partial n_{sub}}\right)^2 \sigma_{n_{sub}}^2 + \left(\frac{\partial \tilde{n}_f}{\partial k_{sub}}\right)^2 \sigma_{k_{sub}}^2 + \left(\frac{\partial \tilde{n}_f}{\partial \lambda}\right)^2 \sigma_\lambda^2 \\ & + 2\left(\frac{\partial \tilde{n}_f}{\partial \psi}\right)\left(\frac{\partial \tilde{n}_f}{\partial \theta_i}\right)\sigma_{\psi\theta_i} + 2\left(\frac{\partial \tilde{n}_f}{\partial \Delta}\right)\left(\frac{\partial \tilde{n}_f}{\partial \theta_i}\right)\sigma_{\Delta\theta_i} \\ & + 2\left(\frac{\partial \tilde{n}_f}{\partial \psi}\right)\left(\frac{\partial \tilde{n}_f}{\partial d_f}\right)\sigma_{\psi d_f} + 2\left(\frac{\partial \tilde{n}_f}{\partial \Delta}\right)\left(\frac{\partial \tilde{n}_f}{\partial d_f}\right)\sigma_{\Delta d_f} + 2\left(\frac{\partial \tilde{n}_f}{\partial d_f}\right)\left(\frac{\partial \tilde{n}_f}{\partial \theta_i}\right)\sigma_{d_f \theta_i} \\ & + 2\left(\frac{\partial \tilde{n}_f}{\partial \psi}\right)\left(\frac{\partial \tilde{n}_f}{\partial n_{sub}}\right)\sigma_{\psi n_{sub}} + 2\left(\frac{\partial \tilde{n}_f}{\partial \Delta}\right)\left(\frac{\partial \tilde{n}_f}{\partial n_{sub}}\right)\sigma_{\Delta n_{sub}} \\ & + 2\left(\frac{\partial \tilde{n}_f}{\partial \psi}\right)\left(\frac{\partial \tilde{n}_f}{\partial k_{sub}}\right)\sigma_{\psi k_{sub}} + 2\left(\frac{\partial \tilde{n}_f}{\partial \Delta}\right)\left(\frac{\partial \tilde{n}_f}{\partial k_{sub}}\right)\sigma_{\Delta k_{sub}} \\ & + 2\left(\frac{\partial \tilde{n}_f}{\partial d_f}\right)\left(\frac{\partial \tilde{n}_f}{\partial n_{sub}}\right)\sigma_{d_f n_{sub}} + 2\left(\frac{\partial \tilde{n}_f}{\partial d_f}\right)\left(\frac{\partial \tilde{n}_f}{\partial k_{sub}}\right)\sigma_{d_f k_{sub}} \\ & + 2\left(\frac{\partial \tilde{n}_f}{\partial \psi}\right)\left(\frac{\partial \tilde{n}_f}{\partial \lambda}\right)\sigma_{\psi\lambda} + 2\left(\frac{\partial \tilde{n}_f}{\partial \Delta}\right)\left(\frac{\partial \tilde{n}_f}{\partial \Delta}\right)\sigma_{\Delta\lambda} + 2\left(\frac{\partial \tilde{n}_f}{\partial d_f}\right)\left(\frac{\partial \tilde{n}_f}{\partial \lambda}\right)\sigma_{d_f \lambda} \end{aligned}} \quad (17)$$

It must be noted that, here, all errors are considered as random noise produced by the measurement. This is not the case for example for an error on $\theta_i$ that should be regarded as a systematic error, a fixed deviation inherent to the measurement configuration and which does not depend on the energy. However, considering only random errors allows us to evaluate their impact on different parts of the energy spectrum.

The thin film thickness that will lead to correct values of $\tilde{n}_f$ with this method depends on the error of the first order approximation. If this error is too large the numerical regression will converge towards incorrect solution.

This point-by-point method allows to explicitly evaluate the accuracy of ellipsometry on the determination of $d_f$, $n_f$ and $k_f$ for the whole spectral range. This information is useful in itself as it can be used for example to evaluate if a spectral feature like a small absorption below the bandgap have a physical origin or if it sits in a range of low accuracy and could then be associated with the error of measurement. The possibility to analyze the error on the whole spectral range is an asset of this method as a similar evaluation is a hard task to do when the analysis is made with a modeled dispersion law like with Tauc-Lorentz oscillators.

The algorithm is available in form of a python code at https://doi.org/10.5281/zenodo.7722620

**III. Theoretical example: very thin film of TiO₂ on Si**

To illustrate the method, we first evaluate it with the theoretical case of a very thin film of $d_f = 1.00\ nm$ of amorphous TiO₂ on silicon substrate for the energy range of 0.77-6.20 eV simulated for an incident angle of $\theta_i = 65°$. TiO₂ is chosen as an example because it has a transparent range in the energy range considered here. Its dispersion law is modeled by a Tauc-Lorentz model (A = 256.08 eV, Br=1.77 eV, E₀ = 4.00 eV, E_g = 3.40eV, $\varepsilon_\infty = 1$) and the optical properties of the silicon substrate are taken from reference [20]. To evaluate the uncertainty of the measurement, we consider here relatively low but realistic errors of $\sigma_\psi = \sigma_\Delta = \sigma_{\theta_i} = 0.01°$ $\sigma_\lambda = 0.1\ nm$ and errors of $\sigma_{n_{sub}} = \sigma_{k_{sub}} = 0.001$.

On Figure 2 (a) the thickness resulting from the McCrakin inversion of this stack is presented together with the calculated thickness from Eq. (9) with their respective error.

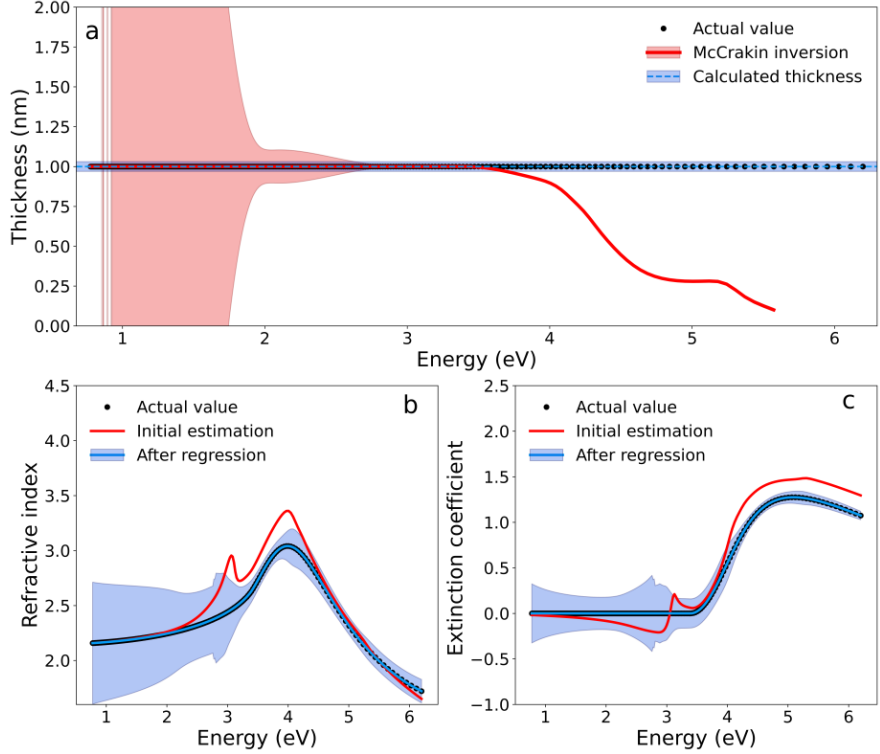

**Figure 2:** (a) Thickness dependence on energy for a thin film of 1 nm (black dots) TiO$_2$ on silicon as determined by a McCrakin inversion method (red line) and calculated from Eq. (9) (blue dashed line). Dispersion law of the refractive index (b) and extinction coefficient (c) of TiO$_2$ (black dots) initial estimation from Eq. (10) (red line) and after the numerical regression (blue line). The colored areas are the calculated errors on the respective values.

As expected, the thickness from the McCrakin inversion shows an energy dependence for values that are above the bandgap of TiO$_2$ ($k_f \neq 0$). Indeed, for this part of the spectrum the hypothesis of a transparent film is not valid, and this range can therefore not be used to determine the thickness. This inversion is therefore already providing interesting information on the dielectric constant of the film that can be used to confirm the values of $\tilde{n}_f$. A region with $k_f \neq 0$ should reflect in a dispersive $d_f$. Below the bandgap the inversion leads to an exact match with the actual value of the thickness. The error distribution with energy also provides valuable information. It shows that the error on $d_f$ exponentially increases with decreasing energy in the low energy range (2.0 – 0.8 eV). This is due to the fact that, for decreasing energy, the difference between $\rho_{sub}$ and $\rho_e$ is decreasing, therefore leading to a higher sensitivity to the measurement parameter errors $\sigma_\psi$, $\sigma_\Delta$, $\sigma_{\theta_i}$, $\sigma_\lambda$ and $\sigma_{n_{sub}}$ and to errors in $\rho_e$. The best energy range to accurately determine $d_f$ is therefore, in this case, 2.0 - 3.4 eV. Using Eq. (9) and the error evaluation in Eq.(8), a precise determination of $d_f = 1.00 \pm 0.03$ nm is achieved.

From the thickness value, a first estimation of $n_f$ and $k_f$ is then made from Eq. (10) and presented on Figure 2 (b) and (c) (red curves). The initial estimation is a relatively good approximation of the dispersion law of TiO$_2$. However, we observe a higher difference between the actual values and estimated ones for a higher energy than for a lower energy range where both values converge. The estimation is based on the first order Taylor expansion, relying on the hypothesis that $\left(\frac{d_f}{\lambda} \ll 1\right)$, hence the error will be increasingly small for decreasing energy (increasing wavelength) as the first order expansion becomes a more accurate approximation. Using this first estimation, the error is then minimized by reducing the error between the measured and modeled $\rho$ values with Eqs. (13)-(16). The values of $n_f$ and $k_f$ after the numerical regression are presented on Figure 2 (b) and (c) (blue curves) with their respective error represented by the colored areas. After the numerical regression the dispersion law of TiO₂ can be perfectly recovered. Regarding the errors on $n_f$ and $k_f$, a relatively low error is observed in both cases in the high energy range (> 3.5 eV) while a large one is observed for the low energy range (< 3.5 eV). Indeed, for decreasing energy, $\delta_\rho$ is decreasing leading to a higher sensitivity to the errors. Around 3 eV, the observed jump in the error is due to the proximity of the two solutions expressed in Eq. (10). Indeed, a small variation of the initial value of the numerical regression will lead to the divergence of the fit towards one solution or the other. Consequently, the error on $n_f$ and $k_f$ is large in the low energy range.

We then considered a thin TiO₂ film of $d_f = 5.00\ nm$ keeping everything else the same. The results are presented on Figure 3.

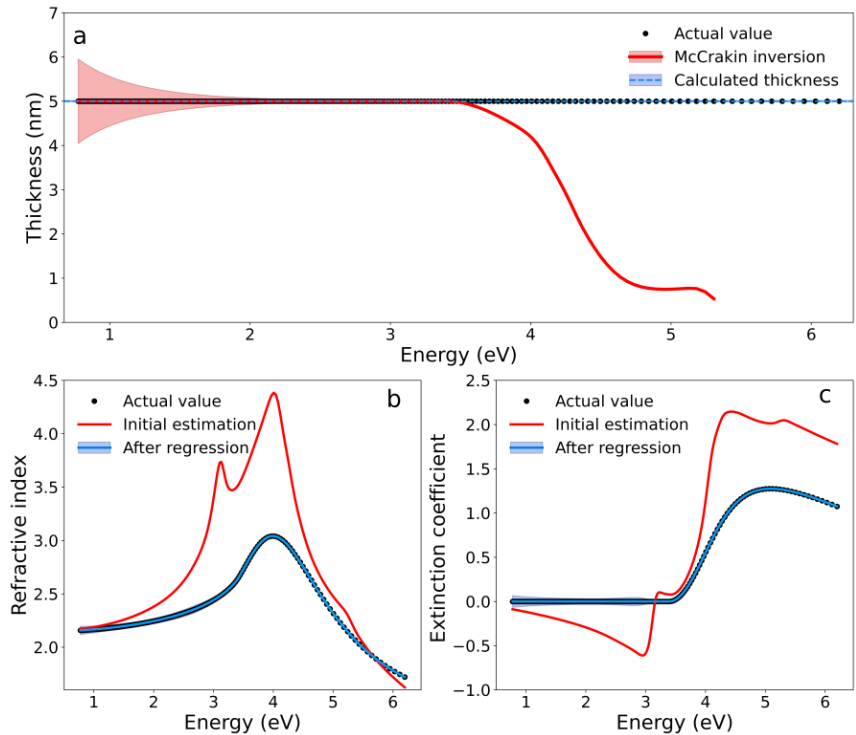

**Figure 3:** (a) Thickness dependence on energy for a thin film of 5.00 nm (black dots) TiO₂ on Silicon as determined by a McCrakin inversion method (red line) and calculated from Eq. (9) (blue dashed line). Dispersion law of the refractive index (b) and extinction coefficient (c) of TiO₂ (black dots) initial estimation from Eq. (10) (red line) and after numerical regression (blue line). The colored areas are the calculated errors on the respective values.

Due to the increased thickness the error on the McCrakin inversion is reduced and an accurate determination can therefore be made in a larger energy range from 1.5 to 3.4 eV (Figure 3 (a)). The calculated thickness from Eq. (9) is $d_f = 5.00 \pm 0.02$ nm. The error on the thickness is much smaller than in the previous case ($d_f = 1.00$ nm) due to a lower dependence on the measurement error as 5 nm leads to a larger difference on $\delta_\rho$. The initial estimation of $n_f$ and $k_f$ is, however, quite different from the actual value (Figure 3 (b) and (c)). This is expected as the first-order Taylor expansion leads to a larger error for thicker films. The regression, however, leads to a very accurate determination of $n_f$ and $k_f$ with a very low error on the considered energy range also due to a higher $\delta_\rho$.

With these two examples, we show that $d_f$, $n_f$ and $k_f$ can be unambiguously determined with the determination of the thickness being made without any assumption on the $n_f$ value. However, in the case of ultrathin films (1 nm), the $n_f$ and $k_f$ values cannot be determined below a given energy (here 3.4 eV) as the error becomes too large. Note that, above a certain film thickness, the method will lead to incorrect values of $\tilde{n}_f$ due to a high error of the first order approximation. The Initial values $n_f$ and $k_f$, would then lead indeed the numerical regression to converge towards one of the incorrect solutions. In this example of a thin film of TiO$_2$ on Si, a thickness larger than 11 nm leads to an incorrect convergence of $n_f$ and $k_f$.

**IV. EXPERIMENTAL EXAMPLE: THIN FILM OF HZO ON SI**

As an experimental example we applied the method to the study of a thin Hf$_{0.5}$Zr$_{0.5}$O$_2$ film grown by atomic layer deposition on RCA cleaned silicon (cf. Experimental details). As a native oxide of SiO$_x$ is present on the surface of Si, the SiO$_x$/Si substrate was measured before deposition to determine $\rho_{sub}$. Then, in order to calculate $\tilde{n}_{sub}$, the pseudo dielectric constant function was calculated[1]. This method allows us to replace a sample that consists of multiple layers by a pseudo dielectric constant that represents the dielectric property of this stack and can thus be considered as the dielectric constant of a new semi-infinite substrate.[21] The errors $\sigma_\psi$, $\sigma_\Delta$ were determined from five measurements of $\psi$ and $\Delta$ on the sample. The error $\sigma_{\theta_i} = 0.1°$ on the angle offset was estimated from five measurements of a 25 nm SiO$_2$ reference sample on Si. The error $\sigma_{\tilde{n}_{sub}}$ on $\tilde{n}_{sub}$ was calculated from the measurements of five RCA cleaned substrates. The resulting calculated thickness $d_f$ is presented on Figure 4 (a).

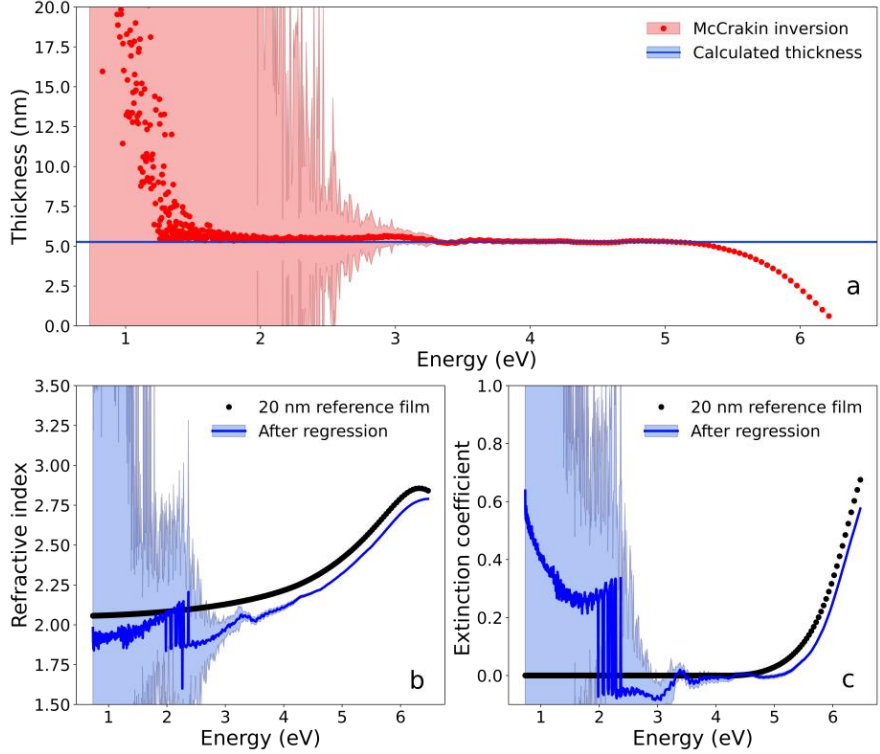

**Figure 4**: (a) Thickness dependence on energy for a ~5 nm HZO film deposited on a RCA cleaned Silicon substrate as determined by a McCrakin inversion method (red line) and calculated from Eq.(9) (blue line) - Dispersion curves of (b) the refractive index and (c) extinction coefficient determined from the proposed method (blue lines) together with the curves for a 20 nm HZO film (black dots). The colored areas are the calculated error on the respective values.

Three regions are observed for the thickness from the McCrakin inversion (Figure 4 (a)). First, there is an energy-dependent region at high energy (~5.2 - 6.5eV). This region corresponds to the non-transparent range of the thin film and cannot be used for the determination of the thickness. Then, a region of constant thickness with low error (~2.5 - 5.2eV) is observed from which the thickness can be accurately determined. A third region is observed at low energy (~0.7 - 2.5eV). In this region, the thickness values show a much larger error and also evolve with both a contribution of a higher scattering and an exponential increase for decreasing energy. As the thickness values are not solely randomly scattered, we can conclude that the observed exponential increase of the thickness comes from systematic errors, such as a constant offset in $\theta_i$, or an error on the dispersion law of $n_{sub}$. A detailed analysis of this region could be done to exploit this artefact to correct for the systematic errors, for example considering adding an offset of 0.1° to the angle of incidence to minimize the thickness evolution with energy. This is, however, outside the scope of this paper. Using Eq. (9) and (8) we calculate the thickness of the film and its corresponding error to be $d_f = 5.27 \pm 0.06$ nm. We can therefore reach a very low uncertainty on the measured thickness thanks to the presence of an energy range with high accuracy in the McCrakin inversion (~2.5-5.2eV).

On Figure 4 (b) and (c) the resulting $n_f$ and $k_f$ values are presented together with the values of a 20 nm thick HZO film. For high energies (> 3.0 eV), we show that both $n_f$ and $k_f$ have dispersion curves similar to their thicker counterpart and that high accuracy ($\sigma_{\tilde{n}_{sub}} \leq 0.01 + 0.002i$) is achieved in both cases. The dispersion law is similar to a dispersion modeled by a Tauc-Lorentz function. Such a dispersion law is characteristic of amorphous materials for which the bandgap is present in the measured spectral region.[1] Note that we obtain such a dispersion law here with this point-by-point method without relying on a model of the dispersion law. The refractive index of the very thin film is lower (2.11 at an energy of 4.0 eV) than the one of the 20 nm film reference (2.21 at an energy of 4.0 eV), which is attributed to a lower density.[1] Using a Bruggeman effective medium approximation, assuming the film is composed of HZO and nanometric air inclusions, we calculate that the very thin HZO film exhibits a density of around 85% that of the 20 nm reference.[22] This is understood by the ALD growth mechanism that tends to generate gaps for the first step of the growth due to steric hindrance.[23] Moreover, from the extinction coefficient a shift towards higher energy of the exponential rise is observed for the thinner film compared to the 20 nm reference film. This shift leads to a higher band gap (5.2 eV) compared to the reference sample (4.9 eV). The increased band gap of the very thin film can be explained by a quantum confinement effect. Indeed, if the dimension of a material is of the same magnitude as the de Broglie wavelength of the electron wave function it will generate a quantum confinement effect. This effect has already been observed during the growth of very thin films.[24]

For energies below ∼ 3 eV the error becomes large, which does not allow to conclude on the optical properties of the HZO thin film. At these energies, similarly to the determination of the thickness, the dispersion laws present errors that are mostly produced from systematic errors of the measurement.

Therefore, at the present state, the proposed method can be applied to accurately measure $d_f$ and $\tilde{n}_f$ of very thin films for energy range higher than ∼ 3 eV. It should be noted that the measurement of a thicker film will widen the energy range where $\tilde{n}_f$ can be retrieved with a high accuracy, as evidenced previously (section III).

**V. CONCLUSION**

In this paper we demonstrated a fully automated method to unambiguously determine the thickness, refractive index and extinction coefficient with high accuracy of a very thin film for energies typically larger than ∼ 3eV. The method is developed for thin films which present a range of transparency and which are deposited on a substrate with known optical properties in the investigated energy range. The method is decomposed in three steps. First, the thickness is estimated from a McCrakin inversion by carefully looking at the energy range with minimal error and thickness dispersion. Second, a first estimation of the refractive index and extinction coefficient is done based on a first order Talyor expansion of $\rho$. Finally, from this initial estimation and the calculated thickness, convergence towards the correct solution of $\tilde{n}_f$ is ensured with a wavelength-by-wavelength numerical regression. We applied the method to a thin HZO film grown by ALD on a silicon substrate and retrieved its optical properties without any model assumption with a high precision in the energy range of 3.0 - 6.7 eV. A high precision (≤0.5 Å) on the determination of the film thickness was also shown.

Calculation of the errors enabled to discriminate physical features from artefacts due to systematic or random errors giving additional information on the sensitivity of the measurement on the whole spectral range. Information on the sensitivity of the measurement in various spectral regions cannot be easily obtained with a standard approach, such as the minimization of the MSE by optimization of the parameters of a Cauchy or Tauc-Lorentz model. With these models it is harder to determine if a spectral feature has a physical origin and should be considered in the model. The proposed method in this work presents a clear advantage in this regard. Exploiting non-physical the exponential dispersions of the thickness, refractive index and extinction coefficient in the low energy range could improve the accuracy of the measurement. This is especially true in the low energy range where sensitivity on the error is the highest.

The need of a transparency range for the thickness determination can be restrictive; it prevents for example from studying the first stages of the growth of a metallic compound. However, the constrain in the first step of the method could be overcome by other methods to disambiguate the thickness from optical properties. One example is the study of the presence of artefacts in the dielectric constant of the substrate.[12,13] The method proposed in this paper is particularly suited for the study of very thin films of oxides, semiconductors or 2D materials, either *ex situ* or *in situ* and in real time particularly during the first stages of the growth. Disambiguating the determination of the thickness from the dielectric properties can genuinely improve the information that can be retrieved from spectroscopic ellipsometry measurements.

## VI. EXPERIMENTAL DETAILS

Prior to thin film deposition, the silicon substrate is cleaned following a standard RCA procedure to remove organic and ionic contaminations and to obtain a defined oxygen-terminated $SiO_x$ surface with the following steps: SC1: 10min, 70-80°C, (5:1:1) H2O + $NH_4OH$ (29% weight) + $H_2O_2$ (30% in solution), HF dip: 15s HF 1%, SC2: 10 min, 70-80°C, (6:1:1) $H_2O$ + 1 HCL (37% weight) + 1 $H_2O_2$ (30% in sol.). The samples are rinsed in $H_2O$ and $N_2$ blow-dried after each step. This standard RCA cleaning process results in a 1.0 nm (±0.1nm) chemical oxide $SiO_x$ layer. Before the deposition, the Si/$SiO_x$ stack is measured by ellipsometry to determine $\rho_{sub}$ and define the pseudo-dielectric function that can be used as the substrate dielectric constant $\tilde{n}_{sub}$. The thin film of HZO is deposited on top of the cleaned Si substrate by atomic layer deposition at 250 °C using tetrakis(ethylmethylamino)zirconium (TEMA-Hf) and tetrakis(ethylmethylamino)zirconium (TEMA-Zr) as precursors and deionized $H_2O$ as oxidant. The depositions are performed using an "Oxford FlexAl" ALD system. The spectroscopic ellipsometry measurements are made with a Woollam M2000 at an incidence angle of 60°, for a wavelength range of 192-1690 nm corresponding to an energy range of 0.73-6.46 eV. The bulk value of $\tilde{n}_{HZO}$ is determined from a Tauc-Lorentz model on a 20 nm thin film deposited with the same conditions.

## ACKNOWLEDGMENTS

The experimental work was performed in the framework of GraFOx II, a Leibniz- ScienceCampus partially funded by the Leibniz Association.

## AUTHOR DECLARATIONS

**Conflict of interest**

The authors have no conflicts to disclose.

**Author contributions**

F. Maudet: Conceptualization (lead); Data Curation (lead); Investigation (lead); Formal Analysis (lead); Methodology (lead); Project Administration (lead); Software (lead); Supervision (lead); Validation (lead); Writing- original draft preparation (lead); Writing- review and editing (equal).

C. V. Dijck: Investigation (supporting); Writing - review & editing (supporting).

M. H. Raza: Investigation (supporting); Writing- review & editing (supporting).

C. Dubourdieu: Conceptualization (supporting); Funding acquisition (lead); Writing- review and editing (lead).

## DATA AVAILABILITY

An archive file with the necessary material of the findings of this study is openly available in Zenodo at https://doi.org/10.5281/zenodo.7722620. The archive contains: a python code of the algorithm of the proposed method, two Jupyter notebooks used to generate the data in the paper alongside with the experimental and reference data used for the analysis.